\documentclass[10pt,conference]{IEEEtran}
\newcommand{\subparagraph}{}
\usepackage{graphicx}
\usepackage{lscape}
\usepackage{amssymb}
\usepackage[letterpaper,top=0.75in,bottom=1.15in,left=0.6in, right=0.6in]{geometry}
\usepackage{amsfonts,amsmath, amssymb, graphics}

\usepackage[compact]{titlesec}

\titlespacing{\section}{0pt}{*0.8}{*0.8}

\titlespacing{\subsection}{0pt}{*0.6}{*0.6}

\begin{document}

\newtheorem{theorem}{Theorem}
\newtheorem{corollary}{Corollary}
\newtheorem{lemma}{Lemma}
\newtheorem{example}{Example}
\newtheorem{definition}{Definition}
\newtheorem{proposition}{Proposition}
\newtheorem{observation}{Observation}
\newtheorem{conjecture}{Conjecture}
\newtheorem{remark}{Remark}
\newtheorem{fact}{Fact}
\newcommand{\N}{\mathbb{N}}
\newcommand{\qed}{\hfill $\diamondsuit$}
\newcommand{\noam}{Noam Presman\ }
\newcommand{\bhat}{Bhattacharyya }
\newcommand{ \etld } {\tilde{\epsilon}}
\newcommand{ \He } {\mathcal{H}}

\title{Design of Non-Binary Quasi-Cyclic LDPC Codes by ACE Optimization}

\author{Alex Bazarsky, Noam Presman and Simon Litsyn \\ School of Electrical Engineering,
Tel Aviv University, Ramat Aviv 69978 Israel \\ e-mail:
\{bazarsky,presmann,litsyn\}@eng.tau.ac.il}

\maketitle

\begin{abstract}
An algorithm for constructing Tanner graphs of non-binary irregular quasi-cyclic LDPC codes is introduced. It employs a new method for selection of edge labels allowing control over the code's non-binary ACE spectrum and resulting in low error-floor. The efficiency of the algorithm is demonstrated by generating good codes of short to moderate length over small fields, outperforming codes generated by the known methods.

\end{abstract}
\section{Introduction}
LDPC codes introduced by Gallager \cite{gal62} are excellent error correcting codes, which are being used in many modern applications.
Non-binary (NB) LDPC codes exhibit better error correcting performance compared to the binary ones \cite{mackay}. As the field size grows, error correcting capability improves, at the price of increasing decoding complexity.

Binary LDPC codes constructed by quasi-cyclic (QC) lifting of a base-graph have a structure that can be utilized in an efficient implementation of both the encoder and the iterative decoder \cite{Mansour,Architecture1,thorpe03}. NB QC codes based on $\alpha$-multiplied circulant permutation matrices have similar properties \cite{ghaffar}.

In LDPC codes, the error-floor is induced by the presence of small combinatorial structures in the Tanner graph (e.g. stopping-sets, trapping-sets, etc.). These structures always contain cycles, therefore manipulating the parameters of the cycles also affects the error-floor. The parameters of importance here are the cycles' length and connectivity, manifested by their approximate cycle extrinsic message degree (ACE). For binary codes, the error-floor can be reduced significantly by removing short cycles having small ACE value from the graph \cite{tian04,dec2008,vuko2008,lit08}. For QC codes, the ACE-constrained construction becomes more computationally efficient by utilizing the relation between cycles in the protograph and their realization in the lifted graph. Based on this idea, Asvadi \textit{et al.}\cite{ban2012} introduced an algorithm for design of irregular binary QC codes with an excellent error-correcting performance.

NB LDPC codes are conventionally constructed by first obtaining a binary mother parity check-matrix $H_b$, and then replacing the non-zero elements of $H_b$ by non-zero values from the field, often referred to as labels.
The label assignment is performed either randomly or intelligently (by meeting some design criteria).

Poulliat \textit{et al.} \cite{dec2008} designed regular NB $(2,d_c)$ codes (cycle codes) by first using  the progressive edge growth (PEG) algorithm \cite{hu2005} to construct $H_b$ having an associated Tanner graph with large girth. Then, they introduced a method for label assignment based on cycle cancelation, resulting in low error-floor codes. Peng and Chen extended this idea for the design of NB QC regular cycle codes \cite{chen2007}. Other design algorithms for NB QC codes were also explored recently \cite{div12} \cite{huang10}.

In this paper, we use a relation between the protograph and the QC lifted graph, to produce good irregular NB QC LDPC codes. Our design involves a new method to select edge labels which constrain the code's NB ACE spectrum, resulting in improved performance.
We demonstrate the efficiency of this algorithm by constructing good codes of short to moderate length over small fields. Such codes are practical due to their moderate decoding complexity. Note that irregular profiles achieve better error correcting performance compared to regular ones for codes over small fields \cite{mackay} \cite{hu2005}, which motivates our ACE based design.

The paper is organized as follows. In Section \ref{sec:Prelim},
we begin by presenting the relevant background and notations that
are used throughout. In Section \ref{code_const}, we present our NB QC code construction method. The performance of codes generated by the method is demonstrated by simulations in Section \ref{sim_results}.

\section{Preliminaries}\label{sec:Prelim}
Throughout we use the following notations. For an integer $n > 0$, let $[n]=\{1,2,\ldots ,n\}$. For two integers $a,b$, the remainder of the division of $a$ by $b$ is denoted by $R_{b}[a]$.
For two vectors $\underline{u},\underline{v}$ of length $\ell$, $\underline{u} \geq \underline{v}$ \textit{iff } $u_i \geq v_i, \,\,\, \forall  i \in[\ell]$.

\setlength{\belowdisplayshortskip}{0pt}
\setlength{\abovedisplayshortskip}{-1\baselineskip plus 3pt}
\subsection{LDPC Codes}
A binary LDPC code of length $n$ is a linear block
code defined by a binary parity-check matrix $H_{m\times n}$.  The code can be
equivalently represented by a bipartite Tanner graph
$G=(V\cup C,E)$, where the set $V$ consists of variable nodes $v_i$,
$i\in[n]$, and the set $C$ consists of check nodes  $c_j$, $j\in
[m]$. An edge connects a variable node $v_i$ to a check node $c_j$
\textit{iff} $H_{j,i}=1$. The degree distribution of the code is represented by two polynomials: $\lambda(x)=\sum_{i=2}^{d_{v}}\lambda_{i}x^{i-1}$ for the variable nodes and $\gamma(x)=\sum_{i=2}^{d_{c}}\gamma_{i}x^{i-1}$ for the check nodes, where $d_{v}$ ($d_c$) is the maximum variable (check) node degree, and $\lambda_{i}$ ($\gamma_i$) is the fraction of edges connected to variable (check) nodes of degree $i$.
If $\lambda(x)=x^{d_v-1}$ and $\gamma(x)=x^{d_c-1}$, the code is called regular $(d_v,d_c)$ LDPC. Otherwise, the code is called irregular LDPC.

An NB LDPC code over $GF(q)$ is defined by a parity-check matrix $H$ with elements from the field ($q=2^r$, $r>1$). The  Tanner graph of the code has labels on its edges which are the corresponding non-zero entries from $H$. For such a code, we define a binary matrix $H_b$ of the same dimensions as $H$, such that each entry in $H_b$ is $1$ \textit{iff} the corresponding entry in $H$ is non-zero. $H_b$ is referred to as the \textit{binary mother matrix} of the code.

\subsection{QC Lifted Codes}\label{QC_codes}

\begin{definition}[Lifted graph]
Let $\hat{G}=(\hat{V} \, \cup \, \hat{C},\hat{E})$ be a Tanner graph. For each $(v,c)\in \hat{E}$ define a permutation $\pi_{(v,c)}$ on the set $[Z]$. To each $v \in \hat{V}$ ($c \in \hat{C}$) we generate a set of $Z$ duplicates $v_i$ ($c_i$), $i \in [Z]$. Then $G=(V \cup C,E)$ is a $Z$-lifted graph of $\hat{G}$, associated with these permutations, if  $V = \{ v_i |
v\in \hat{V}, i\in [Z] \}$, $C=  \{ c_i | c\in \hat{C}, i\in [Z] \}$ and $E
= \left\{ (v_i,c_j) | (v,c)\in \hat{E} \bigwedge
\pi_{(v,c)}(i)=j\right\}$.
\end{definition}

The graph $\hat{G}$ is called a base-graph or a protograph. When all the edge permutations in $G$ are cyclic shifts of $[Z]$, then $G$ is called $Z$-lifted QC graph. In this case, a binary LDPC code associated with $G$, has a compact block-representation of its parity check matrix $H$, based on the parity check matrix $\hat{H}$ associated with the protograph. In this representation, each entry is replaced by a $Z\times Z$ matrix as follows. The zero entries are replaced by zero matrices. Each non-zero entry is replaced by a $(Z,d)$-circulant permutation matrix (CPM), defined below, if a right circular shift of $d$ places is assigned to it.
\begin{definition}[CPM]\label{def:CPM}
A $(Z,d)$-CPM is formed by a circular shift to the right by $d$ places of the columns of the $Z\times Z$ identity matrix.
\end{definition}

In an NB  QC  $Z$-lifted code, each non-zero element of the protograph parity check matrix, corresponding to an edge $e$, is replaced by a $(Z,\lambda_e,\rho_e,d_e)$-multiplied CPM (MCPM), as defined below \cite{ghaffar,chen2007}.
\begin{definition}[MCPM]\label{CPM}
Let $(q-1)|\lambda Z$ and $\alpha$ is a primitive element of $GF(q)$.
A $(Z,\lambda,\rho,d)$-MCPM over $GF(q)$, is a $Z\times Z$ matrix, with underlying binary mother matrix $(Z,d)$-CPM. Furthermore, for each row $i$ of the MCPM $i\in [Z]$, the single non-zero element is $\alpha^{\rho+(i-1)\cdot\lambda}$.

\end{definition}
Note that in a CPM, each row is an $\alpha^{\lambda}$-multiplied circular right shift of the row above it. This is also true for the first row, where the row "above it" is defined to be the last row.

\subsection{Cycles and ACE}

Cycles in the Tanner graph are known to influence the error-floor of iterative decoders (e.g. stopping-sets, trapping-sets, etc.). Important combinatorial characteristics of a cycle are its length and its extrinsic message degree (EMD), which is the number of check-nodes that are connected to the variables of the cycle by only one edge. In this paper, we use for simplicity, the approximate cycle EMD (ACE), defined as $\sum_{v_{i}}(d_{v_{i}}-2)$, where $d_{v}$ is the degree of a node $v$, and the summation is over all the variable-nodes of the cycle. A code with long cycles and large ACE usually exhibits lower error-floor compared to a code with shorter cycles or smaller ACE. This notion motivates the following definition.

\begin{definition}[ACE spectrum]\label{def:Spectrum}
For an LDPC code represented by a Tanner graph $G$, the $\ell$-depth ACE spectrum is $\overline{\boldsymbol\tau}^{(b)}(G)=\left( \tau_{2}^{(b)},\tau_{4}^{(b)},...,\tau_{\ell}^{(b)} \right)$ where $\tau_{i}^{(b)}$, $i \leq \ell$, is the minimum ACE value of any cycle of length $i$ in $G$. $G$ achieves an $\ell$-depth ACE constraint $\hat{\boldsymbol\tau}^{(b)}=\left(\hat{\tau}^{(b)}_{2},\hat{\tau}^{(b)}_{4},...,\hat{\tau}^{(b)}_{\ell} \right)$ if  $\overline{\boldsymbol\tau}^{(b)}(G) \geq \hat{\boldsymbol\tau}^{(b)}$.
\end{definition}
To lower error-floor, it is beneficial to achieve higher ACE spectrum values for cycles of lower length.

We now discuss the relationship between cycles in the protograph $\hat{G}$ and the resultant QC $Z$-lifted graph $G$. Let $\mathcal{C}$ be a cycle in $\hat{G}$ of even length $\ell$, being a sequence of edges $\left\{e_i\right\}_{i=1}^{\ell}$, having ACE $\tau$. Assume that in $G$, for each edge $e_i$ we used a $(Z,d_i)$-CPM, $i \in[\ell]$. The order of $\mathcal{C}$ in $G$ is defined as $O(\mathcal{C}) = Z/\gcd(Z,d)$, where $d=R_Z\left[\sum_{i=0}^{\ell-1}(-1)^{i}d_{i+1}\right]$ is called the \textit{total shift} of the cycle $\mathcal{C}$. It is easy to see that the lifted nodes and edges of $\mathcal{C}$ in $G$ form a union of $\gcd(Z,d)$ cycles, each one of them having length $\ell\cdot O(\mathcal{C})$ and ACE $\tau\cdot O(\mathcal{C})$. Moreover, every cycle in $G$ corresponds to a cycle in $\hat{G}$. Therefore, the ACE spectrum of $G$ can be easily derived from the knowledge about cycles in $\hat{G}$ and the shifts of its edges.

In NB LDPC codes each cycle also has a meaningful algebraic structure, defined by the labels on its edges. A simple and minimal cycle (i.e. that does not contain a cycle being a subset of its nodes), $\mathcal{C}$, of length $\ell$ in an NB Tanner graph $G$, with parity check matrix $H$ can be represented by an $\ell/2\times\ell/2$ matrix denoted $B$. Here, $B$ is the sub-matrix of $H$ with rows and columns that correspond to the check nodes and variable nodes that are in $\mathcal{C}$. Without loss of generality, we can assume that $B$ has the following canonical form. The $i$th row, $i\in \left[\ell/2-1\right]$ is of the form $\left[\mathbf{0}_{i-1}, \beta_{2(i-1)}, \beta_{2(i-1)+1}, \mathbf{0}_{\ell/2-i-1}\right]$ and the last row is $\left[\beta_{\ell-1}, \mathbf{0}_{\ell/2-2},\beta_{\ell-2}\right]$. Here, for $i\geq 0$, $\mathbf{0}_{i}$, is the zero-vector of length $i$ (in case $i=0$, this is the empty vector) and $\left\{\beta_i\right\}_{i=0}^{\ell-1}$ are non-zero elements of $GF(q)$.

In LDPC cycle codes, studied by  Poulliat \textit{et al.} \cite{dec2008}, the variable nodes of every simple and minimal cycle form support of a codeword, unless its corresponding matrix $B$ is full-rank. In a canonical form of $B$, this full-rank condition (FRC) is equivalent to the following

\setlength{\belowdisplayshortskip}{3pt}
\setlength{\abovedisplayshortskip}{-.5\baselineskip plus 3pt}
\begin{equation}\label{FRCS}
(FRC) : \prod_{i=0}^{\ell/2-1}\beta_{2i+1}\neq \prod_{i=0}^{\ell/2-1}\beta_{2i}.
\end{equation}

Therefore, to avoid low-weight codewords in cycle codes (causing high error-floor), Poulliat \textit{et al.}, assign the labels of the NB code, so that cycles of short length fulfill the FRC. A cycle $\mathcal{C}$, that satisfies the FRC is said to be "\textit{canceled}".

We argue that even for general NB irregular codes, assigning labels such that a  cycle $\mathcal{C}$ is canceled, should reduce the probability that the BP iterative decoder fails to converge due to errors in the variables of the cycle. The intuition behind it is that the constraints of the matrix $B$ corresponding to $\mathcal{C}$, imply a local-code on the variables of $\mathcal{C}$ with a single codeword (the zero-codeword) $\textit{iff}$ $\mathcal{C}$ is canceled. Since the iterative decoder is local in its behavior, if $\mathcal{C}$ is not canceled, the decoder could be misled to converge to one of the wrong codewords of the local-code of $\mathcal{C}$. The cycle's extrinsic check nodes may prevent such an erroneous convergence. Having more such extrinsic check nodes should increase the chance to overcome errors in the variables of $\mathcal{C}$. Hence, because label assignment can cancel only a limited number of cycles, it seems reasonable to prefer canceling the shorter ones with low ACE. This notion justifies the next useful definition.

\begin{definition}[NB ACE spectrum]\label{nb_ACE_constraint}
For an NB LDPC code, represented by a Tanner graph $G$, the $\ell$-depth NB ACE spectrum is $\overline{\boldsymbol\tau}^{(nb)}(G)=\left( \tau_{2}^{(nb)},\tau_{4}^{(nb)},...,\tau_{\ell}^{(nb)} \right)$ where $\tau_{i}^{(nb)}$, $i \leq \ell$, is the minimum ACE value of any non-canceled cycle of length $i$ in $G$. $G$ achieves an $\ell$-depth NB ACE spectrum $\hat{\boldsymbol\tau}^{(nb)}=\{\hat{\tau}^{(nb)}_{2},\hat{\tau}^{(nb)}_{4},..,\hat{\tau}^{(nb)}_{\ell}\}$ if  $\overline{\boldsymbol\tau}^{(nb)}(G) \geq \hat{\boldsymbol\tau}^{(nb)}$.
\end{definition}

Peng and Chen \cite{chen2007} showed that for NB QC $Z$-lifted codes, the FRC of the lifted-cycles resulting from a  protograph  cycle $\mathcal{C}$ can be simply expressed through the parameters of the MCPMs assigned to the edges of $\mathcal{C}$.

\begin {theorem}[ \cite{chen2007}]\label{theorem1}
Suppose a simple and minimal cycle $\mathcal{C}$ of length $\ell$ on the protograph is represented by a matrix $B$ in its canonical form. Let $\tilde{B}$ be the QC $Z$-lifted representation of $B$
\begin{equation}\label{cycle_shifts}
\small
\tilde{B}=\left(
  \begin{array}{cccccc}
    P_0 & P_1 & 0 & \ldots & 0 & 0 \\
    0 & P_2 & P_3 & \ldots & 0&0  \\
    0 & 0 & P_4 & \ldots 0  & 0& 0  \\
    \vdots & \vdots & \vdots & \vdots & \vdots & \vdots \\
    0 & 0 & 0 & \ddots &P_{\ell-4}& P_{\ell-3} \\
    P_{\ell-1} & 0 & 0 & \ldots & 0 & P_{\ell-2} \\
  \end{array}
\right),
\normalsize
\end{equation} where $P_i$ is $(Z,\lambda,\rho_i,d_i)$-MCPM. The FRC condition for the cycles induced by $\mathcal{C}$ in the lifted graph (each one of length $O(\mathcal{C})\cdot\ell$) is

\setlength{\belowdisplayshortskip}{0pt}
\setlength{\abovedisplayshortskip}{-1\baselineskip plus 3pt}

\begin{equation}\label{FRCTheo1}
O(\mathcal{C})\cdot\sum_{i=0}^{\ell-1}{(-1)}^i\rho_i \neq 0 \mod (q-1).
\end{equation}
\end{theorem}

Note,  that Peng and Chen required that $(q-1)=Z\cdot\lambda$, however their proof is still valid even in a more general case of $(q-1)|Z\cdot\lambda$. The combinatorial and algebraic connections between the cycles of the protograph and the cycles of the lifted-graph, are a key to the efficient algorithms we present in the next section.

\section{Non-Binary QC ACE Constrained Code Construction}\label{code_const}
In this section, we introduce a code construction algorithm. The inputs to the algorithm are the degree profile of a protograph $\hat{G}$, a lifting order $Z$ and a field size $q$. The algorithm is also given two $\ell$-depth ACE spectrum constraints, $\hat{\boldsymbol\tau}^{(b)}=(\hat{\tau}^{(b)}_2,\ldots,\hat{\tau}^{(b)}_{\ell})$ and $\hat{\boldsymbol\tau}^{(nb)}=(\hat{\tau}^{(nb)}_2,\ldots,\hat{\tau}^{(nb)}_{\ell})$, such that $\hat{\tau}^{(nb)}\geq\hat{\tau}^{(b)}$. The output is $G$, a QC $Z$-lifting of $\hat{G}$ with labels from $GF(q)\setminus\{0\}$,  which is NB ACE constrained by $\hat{\tau}^{(nb)}$ and its binary mother matrix is ACE constrained by  $\hat{\tau}^{(b)}$, if both spectrums are achievable.
The algorithm consists of the following steps: $\\$
\textbf{Step 1:} Construct a good protograph $\hat{G}$ by any protograph selection method (e.g \cite{vuko2008}).  $\\$
\textbf{Step 2:} Construct a QC $Z$-lifted graph of $\hat{G}$, that is $\hat{\boldsymbol\tau}^{(b)}$-ACE constrained, by carefully choosing for each edge of $\hat{G}$ the cyclic shift of its $Z$ copies (see Subsection \ref{bin_const}). This graph is the binary mother matrix of the output $G$.
  $\\$
 \textbf{Step 3:} Assign labels to the edges of the mother matrix, such that the resultant NB labeled graph $G$ is NB ACE constrained by $\hat{\boldsymbol\tau}^{(nb)}$. This label assignment ensures that all the cycles in $G$ that violate  $\hat{\boldsymbol\tau}^{(nb)}$ satisfy the FRC (see Subsection \ref{nb_const}).$\\$

Good achievable constraint vectors $\hat{\boldsymbol\tau}^{(b)}$, $\hat{\boldsymbol\tau}^{(nb)}$ and their depth $\ell$ may be found by the following heuristic search. Find initial constraints $\hat{\boldsymbol\tau}^{(b)}$, $\hat{\boldsymbol\tau}^{(nb)}$ by first running the above algorithm with no constraints, and retrieve the spectra of the resultant graph $G$. Then, attempt to improve the spectra by increasing their depth $\ell$ or increasing their components and rerun the algorithm with the amended spectrum. Repeat this procedure (amending the spectra and rerunning the algorithm) until no further improvement is achieved (i.e. the algorithm fails to find a graph that achieves the constraints). Note that since it is not always possible to determine which spectrum is better (see e.g. \cite[Section IV]{vuko2008}), the designer is advised in these cases, to generate graphs for each of these competing spectra and choose the best one by a simulation. Furthermore, because of the random nature of the algorithm, it is recommended to run the algorithm several times for each set of good parameters, thereby generating different instances of $G$ satisfying the requirements. Here, again, the best instance, may be chosen by a simulation.
\subsection{Construction of the Binary Mother Matrix}\label{bin_const}
We now describe an algorithm that finds a QC binary code that satisfies certain ACE spectrum constraints. The inputs to the algorithm are a protograph $\hat{G}$, a lifting factor $Z$ and an ACE spectrum constraint vector $\hat{\boldsymbol\tau}^{(b)}=\left(\hat{\tau}^{(b)}_{2},\hat{\tau}^{(b)}_{4},..,\hat{\tau}^{(b)}_{\ell}\right)$. The algorithm searches for a $QC$ $Z$-lifted code with Tanner graph $G$ which achieves  $\hat{\boldsymbol\tau}^{(b)}$. $G$ is defined by assigning a cyclic shift $d_e \in [Z]$ to each edge $e$ of $\hat{G}$.

We begin by a preliminary step in which we find all the problematic cycles of $\hat{G}$ which violate $\hat{\boldsymbol\tau}^{(b)}$. Denote this set of problematic cycles by $S$. The lifted versions of the other cycles of $\hat{G}$ will satisfy the $\hat{\boldsymbol\tau}^{(b)}$ constraint for any choice of shifts.
Next, for each edge $e$ of $\hat{G}$ we enumerate the problematic cycles which include $e$. We arbitrarily choose initial assignments of $d_e$ for each edge $e$ of $\hat{G}$ (it is recommended to draw these assignments uniformly at random).

The generation of $G$ is iterative. In each iteration, we scan all the edges of $\hat{G}$ in an arbitrary order. For each edge $e$, we choose a shift $d_e \in [Z]$ that minimizes the number of cycles in $S$ which still violate the $\hat{\boldsymbol\tau}^{(b)}$ constraint. If by the end of the iteration, all the lifted versions of the cycles in $S$ satisfy $\hat{\boldsymbol\tau}^{(b)}$, the algorithm outputs $G$, otherwise, another iteration may be initiated. Note that there is no guarantee that the algorithm finds $G$ that achieves $\hat{\boldsymbol\tau}^{(b)}$ even if such $G$ exists. If after a predefined number of iterations $G$ is not found, the designer may consider choosing a different initial assignment of the shifts $d_e$ or changing the order in which the edges are visited, and repeat the iterative part of the algorithm. Our experience shows that usually when $G$ exists, it is found after a small number of iterations.

Our method differs from the algorithm of Asvadi \textit{et al.} \cite{ban2012,asvadi2011} in the following aspects.
In \cite[Algorithm 1]{asvadi2011}, the cycles in $S$ are sequentially scanned and for each cycle, the shifts $d_e$ of the "unshifted edges" are assigned whereas in our method, the objects being treated are the edges.
Furthermore, we allow reassignment of $d_e$ in case the ACE spectrum constraint was not satisfied after the first iteration. Our experience indicates that in many cases our proposed algorithm finds $G$ faster than \cite[Algorithm 1]{asvadi2011}.
Peng and Chen \cite{chen2007}, suggested to find the binary lifted graph $G$ of cycle codes, having girth $\ell$. Note that this is equivalent to having ACE spectrum constraints $\hat{\boldsymbol\tau}^{(b)}$ such that $\hat{\tau}_i^{(b)}=\infty$ for all $i \leq \ell$.
Such a spectrum can be achieved  only for relatively small $\ell$. As a result,  longer cycles with low ACE are ignored. Our results indicate that these cycles have an impact on the code's performance in the error-floor region.

\subsection{Non-Binary ACE Constrained Label Assignment}\label{nb_const}
We now describe the label assignment algorithm to the entries of the binary mother matrix. The inputs to the algorithm are the QC $Z$-lifted graph $G$ (associated with the binary mother matrix) expressed as the underlying protograph $\hat{G}$ and the selected shifts of its edges.  Additional inputs are the NB ACE constraint vector $\hat{\boldsymbol\tau}^{(nb)}$ and the field size $q$. The output of the algorithm is an NB QC code over $GF(q)$ that satisfies $\hat{\boldsymbol\tau}^{(nb)}$.

The structure of the algorithm is very similar to the one presented in the previous subsection, therefore we only highlight here the differences between them.
We begin by enumerating the set $S$ of problematic cycles in $\hat{G}$. This time a cycle is problematic if its $Z$-lift in $G$ violates $\hat{\boldsymbol\tau}^{(nb)}$. For each edge $e$ in $\hat{G}$ we choose an initial label $\rho_e$ (preferably at random). We then run the iterative part of the algorithm from Subsection \ref{bin_const} in which we assign labels $\rho_e$ (instead of shifts). Note that, in this case, a problematic cycle violates  $\hat{\boldsymbol\tau}^{(nb)}$ if its labels do not satisfy (\ref{FRCTheo1}).

The algorithm we described can be seen as a generalization of the method suggested by Peng and Chen \cite{chen2007} for construction of NB QC lifted graphs of cycle codes. In their algorithm, all cycles up to length $\ell$ are canceled, which is equivalent to using our algorithm with $\hat{\boldsymbol\tau}^{(nb)}$ such that $\hat{\tau}^{(nb)}_i=\infty$ for all $i \leq\ell$.
Furthermore, we allow a more flexible choice of the lifting order $Z$ and the CPM parameter $\lambda$, by only requiring that $(q-1)|Z\lambda$ (Peng and Chen's requirement is that $Z|(q-1)$ and $\lambda=(q-1)/Z$).

\section{Simulation Results}\label{sim_results}
In this section, we compare the error correcting performance of codes generated by the following techniques:
\textbf{PEG} - PEG generated mother binary matrix of an irregular code with randomly selected labels \cite{hu2005}.
\textbf{REG} - PEG generated mother binary matrix of a regular cycle code $(d_v=2,d_c=4)$ with selective choice of NB labels using the FRC \cite{dec2008}.
\textbf{GIRTH} - the QC construction presented by Peng and Chen \cite{chen2007} modified to produce irregular codes, where only the girth requirements are taken into account.
\textbf{ACENB} - our algorithm from Section \ref{code_const}.
\textbf{ACEB} - our algorithm \textbf{ACENB} in which Step 3 is replaced by random edge label assignment.

All the codes are simulated over a memoryless binary input AWGN channel using  the BPSK modulation and decoded by the iterative NB belief-propagation (BP) algorithm.
The maximum number of BP flooding iterations is fixed to $80$.
All the generated codes are of rate $1/2$, and their ensemble properties are summarized in Table \ref{deg_table}. The degree distributions are selected according to Hu \textit{et al.} \cite{hu2005}. Note that $\lambda(x)$ and $\gamma(x)$ are degrees profiles of the protograph and $Z$ is the lifting order, resulting in an ensemble of QC codes of the specified length in bits. For each ensemble, we use a single protograph matrix (Step 1 in Section \ref{code_const}) throughout the various generation methods.

In Table \ref{ace_table}, the achieved ACE spectra are summarized. For the codes, generated by \textbf{ACENB} and \textbf{GIRTH}, two spectra are provided. The first one is the spectrum achieved by the binary mother matrix (Step 2 in our code construction). The second one is the NB ACE spectrum achieved by the edge labels assignment (Step 3 in our code construction). For the codes generated by \textbf{ACEB}, only one spectrum is provided, since the random label assignment does not take into account any NB ACE spectrum requirement. The achieved ACE spectrum of the codes generated by \textbf{GIRTH}, depends only on the codes' girth, while the NB ACE spectrum depends only on the canceled cycles' length. Note that the achievable ACE spectrum values grow with the code length.

\begin {table}[ht]
\fontsize{7}{8}
\begin{tabular}{|c|c|c|c|c|c|}
  \hline
   \# & Field & Length  & $\lambda(x)$ & \parbox[c][0.5cm][c]{0.6cm}{$\gamma(x)$} & $Z$ \\ 
   & & [bits]& & & \\ \hline
  1 & GF(16) & 504 & \parbox[c][0.5cm][c]{3cm}{$0.588x$+$0.176x^2$+$0.235x^3$} & \parbox[c][0.5cm][c]{2cm}{$0.118x^3$+$0.882x^4$} & 9 \\ \hline
  2 & GF(8) & 1008 & \parbox[c][0.5cm][c]{3cm}{$0.487x$+$0.22x^2$+$0.292x^3$} & \parbox[c][0.5cm][c]{2cm}{$0.853x^4$+$0.146x^5$} & 21  \\ \hline
  3 & GF(16) & 1008 & \parbox[c][0.5cm][c]{3cm}{$0.588x$+$0.176x^2$+$0.235x^3$} & \parbox[c][0.5cm][c]{2cm}{$0.118x^3$+$0.882x^4$} & 18 \\ \hline
   4 & GF(16) & 1512 & \parbox[c][0.5cm][c]{3cm}{$0.588x$+$0.176x^2$+$0.235x^3$} & \parbox[c][0.5cm][c]{2cm}{$0.118x^3$+$0.882x^4$} & 27 \\ \hline
\end{tabular}
$\\$
\caption{Ensemble properties of the generated codes }
\label{deg_table}
\vskip -0.7cm
\end {table}

\begin {table}[ht]
\fontsize{7}{8}
\begin{tabular}{|c|c|c|c|}
                   \hline

                   \# & \parbox[l][0.2cm][l]{0.7cm}{\textbf{GIRTH}} & \parbox[c][0.6cm][c]{0.7cm}{\textbf{ACEB}} & \parbox[c][0.6cm][c]{1cm}{\textbf{ACENB}} \\ \hline
                  1   & \parbox[c][0.7cm][c]{2.3cm}{$\tau_i=\infty , i \leq 6$ \\ $\tau_i=\infty , i \leq 8$}  & \parbox[l][0.3cm][l]{2.2cm}{$(\infty,\infty,\infty,4)$} & \parbox[l][0.6cm][l]{2.75cm}{$(\infty,\infty,\infty,4)$ \\ $(\infty,\infty,\infty,\infty,\infty,4)$}\\ \hline
                   2 &\parbox[l][0.7cm][c]{2.3cm}{$\tau_i=\infty , i \leq 6$ \\ $\tau_i=\infty , i \leq 8$} & \parbox[l][0.3cm][l]{2.2cm}{$(\infty,\infty,\infty,5,2)$} & \parbox[l][0.6cm][l]{2.75cm}{$(\infty,\infty,\infty,6,2)\\ (\infty,\infty,\infty,\infty,6,2)$}  \\ \hline
                   3 & \parbox[l][0.7cm][c]{2.3cm}{$\tau_i=\infty , i \leq 8$ \\ $\tau_i=\infty , i \leq 10$} & \parbox[l][0.3cm][l]{2.2cm}{$(\infty,\infty,\infty,\infty,3,1)$} & \parbox[l][0.6cm][l]{2.75cm}{$(\infty,\infty,\infty,\infty,3,1)\\ (\infty,\infty,\infty,\infty,\infty,6,2)$}  \\
                   \hline
                  4 & \parbox[l][0.7cm][c]{2.3cm}{$\tau_i=\infty , i \leq 8$ \\ $\tau_i=\infty , i \leq 10$} & \parbox[l][0.3cm][l]{2.2cm}{$(\infty,\infty,\infty,\infty,4,2)$} & \parbox[l][0.6cm][l]{2.75cm}{$(\infty,\infty,\infty,\infty,4,2)\\ (\infty,\infty,\infty,\infty,\infty,9,3)$}  \\
                   \hline
\end{tabular}
$\\$
\caption{ACE spectra achieved by each of the generated codes. For columns \textbf{ACEB} and \textbf{ACENB}, the spectrum's format is $(\tau_2,\tau_4,\tau_6,\ldots)$.}
\label{ace_table}
\vskip -0.5cm
\end {table}

\begin{figure}
\hspace*{-0.3in}
  \includegraphics[scale = 0.52]{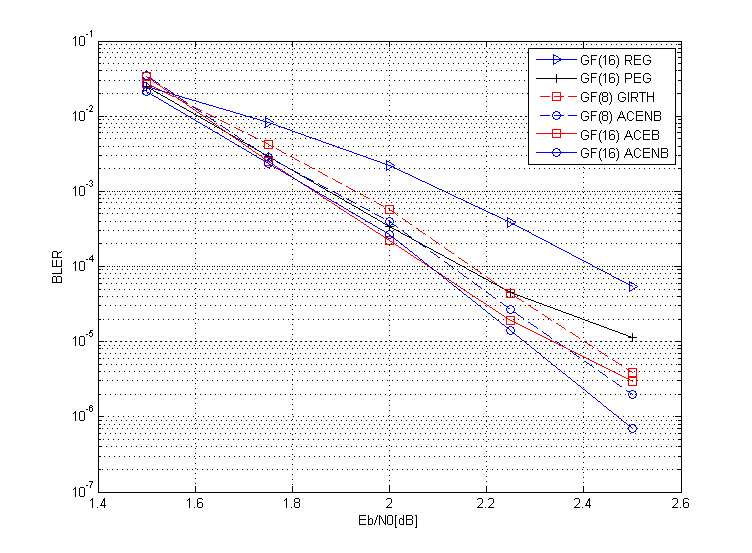}
  \vskip -0.4cm
  \caption{Simulation of codes having length $n_b=1008$ and rate 1/2, constructed by various methods over GF(8) and GF(16)  }\label{fig:1}

\end{figure}

In Figure \ref{fig:1}, the block error rate (BLER) curves of codes from ensembles \#2 and \#3 from Table \ref{deg_table} are depicted.
As expected, the regular code generated by \textbf{REG} is inferior to the other irregular codes. Furthermore, even though the code generated by \textbf{PEG} is not constrained by the QC requirement, its performance in the high SNR region is worse than our QC codes.
 For each ensemble, it is evident that the code generated by \textbf{ACENB} outperforms the codes generated by the other methods in the high SNR region. Since the codes generated by \textbf{ACEB} and \textbf{ACENB} have the same binary mother matrix, the error curves highlight the advantage of applying Step 3 in \textbf{ACENB} instead of random label assignment. Also, with the increase of the field size, we see an improvement in the high SNR region.

\begin{figure}
\hspace*{-0.27in}
  \includegraphics[scale = 0.58]{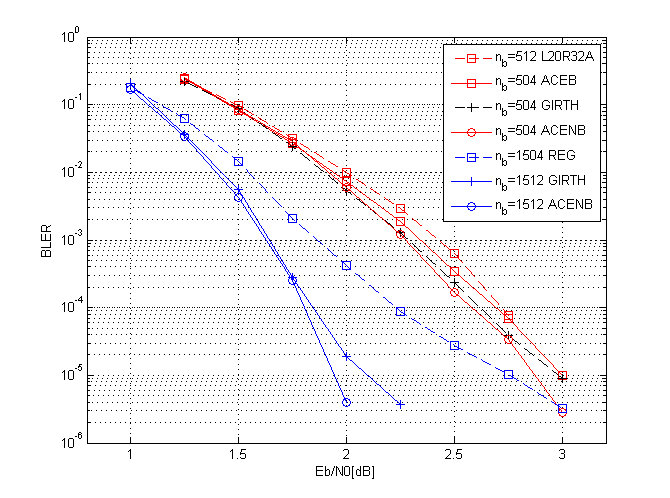}
   \vskip -0.4cm
  \caption{Simulation of rate 1/2 codes having lengths around $n_b=512$ and $n_b=1504$ over GF(16)   }\label{fig:2}

\end{figure}

In Figure \ref{fig:2}, the BLER curves of codes from ensembles \#1 and \#4 from Table \ref{deg_table} are depicted.
 We also give as a reference the BLER curve of \textbf{L20R32A}, a QC code recently generated by Chang \textit{et al.} \cite[Figure 10]{div12}.
Note that, the performance of the code generated by \textbf{REG} agrees with the results of Poulliat \textit{et al.} \cite[Figure 4]{dec2008}, which is simulated using $1000$ as the maximum number of BP iterations. To have a fair comparison with the latter code, we use the same limitation on the number of iterations for the codes from ensemble \#4.
The curves indicate that codes generated by \textbf{ACENB} outperform the other codes with similar lengths in the high SNR region.
Furthermore, the comparison of the curves corresponding to \textbf{ACENB} and \textbf{GIRTH} indicates the advantage of ACE driven construction of NB codes.
\section{Summary and Conclusions}\label{conclusions}

We presented an algorithm to design the Tanner graph of NB QC LDPC codes using ACE constraints for both the generation of the binary mother matrix and the selection of NB labels. Our simulation results indicate that codes generated by this method outperform codes generated by known methods, for different small field sizes and code lengths.

Our method is composed of three separate steps. It is an open question whether combining could be beneficial. This is a matter for future research.

\bibliographystyle{IEEEtran}

\bibliography{IEEEabrv,ldpcbase}

% Generated by IEEEtran.bst, version: 1.13 (2008/09/30)
\begin{thebibliography}{10}
\providecommand{\url}[1]{#1}
\csname url@samestyle\endcsname
\providecommand{\newblock}{\relax}
\providecommand{\bibinfo}[2]{#2}
\providecommand{\BIBentrySTDinterwordspacing}{\spaceskip=0pt\relax}
\providecommand{\BIBentryALTinterwordstretchfactor}{4}
\providecommand{\BIBentryALTinterwordspacing}{\spaceskip=\fontdimen2\font plus
\BIBentryALTinterwordstretchfactor\fontdimen3\font minus
  \fontdimen4\font\relax}
\providecommand{\BIBforeignlanguage}[2]{{%
\expandafter\ifx\csname l@#1\endcsname\relax
\typeout{** WARNING: IEEEtran.bst: No hyphenation pattern has been}%
\typeout{** loaded for the language `#1'. Using the pattern for}%
\typeout{** the default language instead.}%
\else
\language=\csname l@#1\endcsname
\fi
#2}}
\providecommand{\BIBdecl}{\relax}
\BIBdecl

\bibitem{gal62}
R.~Gallager, ``Low-density parity-check codes,'' \emph{Information Theory, IRE
  Transactions on}, vol.~8, no.~1, pp. 21 --28, january 1962.

\bibitem{mackay}
M.~Davey and D.~MacKay, ``Low-density parity check codes over {GF}(q),''
  \emph{Communications Letters, IEEE}, vol.~2, no.~6, pp. 165--167, 1998.

\bibitem{Mansour}
M.~Mansour and N.~Shanbhag, ``High-throughput {LDPC} decoders,'' \emph{IEEE
  Trans. on Very Large Scale Integration Systems}, vol.~11, no.~6, pp.
  976--996, 2003.

\bibitem{Architecture1}
V.~Novichkov, H.~Jin, and T.~Richardson, ``Programmable vector processor for
  irregular {LDPC} codes,'' in \emph{38th Annual Conf. on Info. Sciences and
  Systems}, March 2004.

\bibitem{thorpe03}
J.~Thorpe, ``Low-density parity-check (ldpc) codes constructed from
  protographs,'' \emph{IPN Progress Report, Tech. Rep. 42-154}, 2003.

\bibitem{ghaffar}
L.~Zeng, L.~Lan, Y.~Tai, S.~Song, S.~Lin, and K.~Abdel-Ghaffar, ``Constructions
  of nonbinary quasi-cyclic {LDPC} codes: A finite field approach,''
  \emph{{IEEE} Trans. Commun.}, vol.~56, no.~4, pp. 545 --554, april 2008.

\bibitem{tian04}
T.~Tian, C.~Jones, J.~Villasenor, and R.~Wesel, ``Selective avoidance of cycles
  in irregular {LDPC} code construction,'' \emph{{IEEE} Trans. Commun.},
  vol.~52, no.~8, pp. 1242 -- 1247, aug. 2004.

\bibitem{dec2008}
C.~Poulliat, M.~Fossorier, and D.~Declercq, ``Design of regular
  (2,$d_c$)-{LDPC} codes over {GF}(q) using their binary images,'' \emph{{IEEE}
  Trans. Commun.}, vol.~56, no.~10, pp. 1626 --1635, october 2008.

\bibitem{vuko2008}
D.~Vukobratovic and V.~Senk, ``Generalized {ACE} constrained progressive
  edge-growth {LDPC} code design,'' \emph{{IEEE} Commun. Lett.}, vol.~12,
  no.~1, pp. 32 --34, january 2008.

\bibitem{lit08}
E.~Sharon and S.~Litsyn, ``Constructing {LDPC} codes by error minimization
  progressive edge growth,'' \emph{{IEEE} Trans. Commun.}, vol.~56, no.~3, pp.
  359--368, 2008.

\bibitem{ban2012}
R.~Asvadi, A.~Banihashemi, and M.~Ahmadian-Attari, ``Design of finite-length
  irregular protograph codes with low error floors over the binary-input {AWGN}
  channel using cyclic liftings,'' \emph{{IEEE} Trans. Commun.}, vol.~60,
  no.~4, pp. 902 --907, april 2012.

\bibitem{hu2005}
X.-Y. Hu, E.~Eleftheriou, and D.~Arnold, ``Regular and irregular progressive
  edge-growth tanner graphs,'' \emph{{IEEE} Trans. Inf. Theory}, vol.~51,
  no.~1, pp. 386 --398, jan. 2005.

\bibitem{chen2007}
R.-H. Peng and R.-R. Chen, ``Design of nonbinary quasi-cyclic {LDPC} cycle
  codes,'' in \emph{Information Theory Workshop, 2007. ITW '07. IEEE}, sept.
  2007, pp. 13 --18.

\bibitem{div12}
B.-Y. Chang, D.~Divsalar, and L.~Dolecek, ``Non-binary protograph-based {LDPC}
  codes for short block-lengths,'' in \emph{Information Theory Workshop (ITW),
  2012 IEEE}, Sept., pp. 282--286.

\bibitem{huang10}
J.~Huang, L.~Liu, W.~Zhou, and S.~Zhou, ``Large-girth nonbinary {QC-LDPC} codes
  of various lengths,'' \emph{{IEEE} Trans. Commun.}, vol.~58, no.~12, pp.
  3436--3447, Dec. 2010.

\bibitem{asvadi2011}
R.~Asvadi, A.~Banihashemi, and M.~Ahmadian-Attari, ``Lowering the error floor
  of {LDPC} codes using cyclic liftings,'' \emph{{IEEE} Trans. Inf. Theory},
  vol.~57, no.~4, pp. 2213 --2224, april 2011.

\end{thebibliography}

\end{document}